\documentstyle[preprint,revtex]{aps}
\begin{document}
\hfill\vbox{\hbox{\bf NUHEP-TH-93-23}\hbox{July 1993}\hbox{revised Oct 1993}
}\par
\thispagestyle{empty}
\begin{title}
{\bf Photoproduction of $WH$ signal at electron-proton Colliders}
\end{title}
\author{Kingman~Cheung}
\begin{instit}
Dept. of Physics \& Astronomy, Northwestern University, Evanston,
Illinois 60208, USA\\
\end{instit}
\begin{abstract}
\nonum
\section{Abstract}
We present the photoproduction of an intermediate mass Higgs  (IMH)
boson associated with a  $W$ boson at the future
electron-proton colliders using bremsstrahlung photon beam or laser
backscattered photon beam.
With bremsstrahlung photon beam the  search for the IMH boson is unfavorable
because of the small signal rate.
But with laser photon beam  the search is viable due to a much larger rate,
and provided that the $B$-identification is efficient and $m(b\bar b)$
measurement has a good resolution.
\end{abstract}

\begin{center}
{\it (appear in Phys. Lett. B319, 244)}
\end{center}

\newpage
\section{Introduction}
\label{intro}

Standard Model (SM) Higgs boson   has been searched at LEP and the negative
results put a
lower bound of about 60 GeV \cite{mhlep} on the Higgs-boson mass.
With the highest
center-of-mass energy of LEP~II the search can push the limit to about
90~GeV.  The discovery of a heavy Higgs boson, on the other hand, should be
viable using the $H\to ZZ,\,WW$ decay channels at the future hadronic
supercolliders \cite{ZZWW}.  But so far there are still controversies on
identifying the intermediate mass Higgs (IMH) boson at the hadronic
supercolliders.   On the other hand, the whole intermediate mass range
should be covered at the future linear $e^+e^-$
colliders of $\sqrt{s}=300-500$~GeV \cite{imh}.
Besides, the linear $e^+e^-$ colliders
operating in $\gamma\gamma$ or $e\gamma$ modes, where the photon beams are
realized by the laser backscattering method \cite{teln},
have also been shown to  be possible
for the IMH discovery \cite{bowser,WH}.  The feasibility for IMH discovery
has also been explored  at $ep$
colliders via $ep\to \nu HX$ production channel \cite{zepp}, however, a
detector of high performance is necessary  for reasonable
signal-to-background ratios.
Another possibility will be the associated photoproduction of an IMH boson
 with a $W$ boson at $ep$ colliders via the subprocess
\begin{equation}
\label{one}
  q\gamma \to q' W^\pm  H\,,
\end{equation}
where the initial photon is obtained by the bremsstrahlung off the
incoming electron or by the laser backscattering method \cite{teln}.
This production channel has the advantage of an additional $W$ boson which can
be tagged by its leptonic or hadronic decays
to eliminate the QCD backgrounds.
In this paper  we will investigate this possibility and
will show that this channel can only give a small
cross section for the bremsstrahlung photons but a much larger
cross section for the laser backscattered photons at the future $ep$
colliders.
We will also discuss the corresponding backgrounds in the search of
the IMH boson at the LEP$\times$LHC.

\section{Photoproduction}

HERA is at present the only $ep$ collider running. One of its goals is to
provide data for accurate determination of the proton structure function at
the small $x$ region.  Another $ep$ collider \cite{data}
being contemplated is the combination of the LEP and the LHC (LEP$\times$LHC),
in which an 60 GeV electron beam collides with a
7.7 TeV proton beam, and the center-of-mass energy $\sqrt{s}$ of the
collision is 1.36 TeV.  The luminosity under consideration is about $2.8\times
10^{32}$~cm$^{-2}$~s$^{-1}$.  If we assume a 30\% duty cycle (same as the SSC),
then in one year of running it can accumulate about 2.65 fb$^{-1}$.  For the
following discussion we will assume a yearly luminosity of 3 fb$^{-1}$.

Photoproduction \cite{photo} at $ep$ (HERA, LHC$\times$LEP)  colliders is a
part of the  neutral-current (NC)
scattering, $ep\to eX$, with the exchange photon being almost on-shell
$Q^2\approx 0$. In this type of events the scattered electron is tagged in the
direction of the incoming electron beam within a certain angle,
and its energy is measured.  Using
this scattered electron as a trigger, a lot of charged-current (CC)
backgrounds can be eliminated.  Nevertheless, there are still large
backgrounds coming from the NC photoproduction of jets.

We will use the Weizs\"{a}cker-Williams approximation (WWA) of the form
\begin{equation}
f^{\rm WWA}_{\gamma/e}(x) = \frac{\alpha}{2 \pi} \; \frac{1 + (1-x)^2}{x}
 \log \frac{Q_{\rm max}^{2 \rm eff} } {Q_{\rm min}^2}\,,
\end{equation}
for the luminosity function of the bremsstrahlung photons.
The $Q_{\rm min}^2= m_e^2 x^2 /(1-x)$ is determined by the kinematic limit.
The $Q^{2 \rm eff}_{\rm max}$  is expected to be less than
the maximum virtuality of the photon, and is taken  to be one quarter
the square of the center-of-mass energy of the $ep$ system.
For the proton structure function we will use the HMRS (set b) \cite{HMRS}
with the scale $\mu^2 = M^2(WH) + p^2_T(WH)$, the transverse mass of the $WH$
pair.  The effective $\gamma p$ luminosity is equal to the $ep$ luminosity in
the approximation of one photon bremsstrahlung.

On the other hand,
the laser backscattered photons are obtained by directing a low-energy but
intense laser beam almost head-to-head to the incident electrons.  The
resulting photon beam carries most of the energy of the electron beam and
therefore has the advantage that its spectrum is much harder than that of the
bremsstrahlung photons.  The spectrum for the  unpolarized beam  is given by
\cite{teln}
\begin{mathletters}
\begin{eqnarray}
\label{f-lasera}
f^{\rm laser}_{\gamma/e} (x)& = & \frac{1}{D(\xi)} \left[ 1-x +\frac{1}{1-x}
-\frac{4x}{\xi(1-x)} + \frac{4x^2}{\xi^2 (1-x)^2} \right] \,, \\
\label{f-laserb}
D(\xi) & = & (1-\frac{4}{\xi} -\frac{8}{\xi^2}) \ln(1+\xi) + \frac{1}{2} +
\frac{8}{\xi} - \frac{1}{2(1+\xi)^2}\,,
\end{eqnarray}
\end{mathletters}
where $\xi = 4E_e\omega_0/m_e^2$,  and $\omega_0$ is the energy of
the incident laser photon.   $\xi$ is chosen to be about 4.8 and $x_{\rm max}$
is then about 0.83.  In this method, if the converted electron can be removed
efficiently the charged-current backgrounds will be eliminated, and resulting
in  pure $\gamma p$ collision.
The effective $\gamma p$ luminosity depends on how efficient the conversion
$e\to \gamma$ is.  According to Ref.~\cite{teln}, the effective $\gamma\gamma$
luminosity can be achieved higher than the original $e^+e^-$ collider,
but for $e\gamma$ collision the effective $e\gamma$ luminosity
was shown to be limited \cite{teln}.
So far the use of laser backscattering method to
convert $e\to\gamma$ in $ep$ environment has not been studied.
For the calculation we are going to present we assume
that the effective $\gamma p$ luminosity is the same as the $ep$ luminosity.

The contributing Feynman diagrams for the subprocess
\begin{equation}
d(p_1) \;\gamma(p_2) \to  u(q_1)\; W^-(k_1)\; H(k_2)\;,
\end{equation}
are  shown  in Fig.~\ref{feyn}.  The helicity amplitudes for the matrix
elements have been given for a similar process $e^-\gamma\to W^-H\nu$ in
Ref.~\cite{WH}.  Simple adaptation from those formulas by changing the
corresponding photon coupling to quarks should be made.  The matrix elements
for the charge-conjugate process
\begin{equation}
u(p_1) \;\gamma(p_2) \to  d(q_1)\; W^+(k_1)\; H(k_2)\;
\end{equation}
can be obtained simply by charge-conjugation.
The total cross section is obtained by convoluting the subprocess cross
section with the proton structure
function and the luminosity function for the bremsstrahlung photons or for
the laser backscattered photons.

\section{Results}

We used the input parameters of $M_Z=91.175$~GeV and $\sin^2\theta_{\rm W}=
0.23$. The total cross section of the photoproduction of $ep\to WHX$ using
the WWA  for the bremsstrahlung photons versus the
center-of-mass energies of the $ep$ system for a range of Higgs masses from 60
-- 140~GeV is shown in Fig.~\ref{cross}.  The cross section starts off with a
very small value, and increases quite rapidly around $\sqrt{s}=1-2$~TeV
because the phase space factor to produce $WH$ final state is very limited
at small $\sqrt{s}$  but  increases
favorably with $\sqrt{s}$.  At $\sqrt{s}=3$~TeV the cross section for
$m_H=60$~GeV can reach 32 fb.
On the other hand, using the laser backscattered photons the production cross
section is more than an order of magnitude larger, as shown in
Fig.~\ref{cross-l}.  Therefore, the plausibility of using the $WH$ channel
to search for the IMH boson improves substantially.
For the following discussions we will concentrate on the LEP$\times$LHC
of $\sqrt{s}=1.36$~TeV.  The production rates for $\sqrt{s}=1.36$~TeV using the
bremsstrahlung photons and laser photons are listed in Table~\ref{table}.

\section{Discussions}

So far the studies on the IMH search at $ep$ collision are not so ideal
because  of the presence of huge QCD backgrounds from NC and CC scatterings
\cite{zepp}.
In our study we can tag on the charged lepton  from the leptonic $W$ decay or
on the pair of jets from the hadronic $W$ decay.
This can help suppressing the QCD backgrounds.
The dominant decay of the IMH  is the $b\bar b$ mode, which is chosen to
maximize the  number of signal events. Therefore, in the final state we have
\begin{equation}
ep \to WHX \to \ell \nu (b\bar b) X\,,
\end{equation}
or
\begin{equation}
ep \to WHX \to (jj) (b\bar b) X\,.
\end{equation}
The irreducible backgrounds from $ep\to WZX \to W(b\bar b)X$ and $ep\to
\bar t bX \to b\bar b WX$ are always present, no matter which decay mode of the
$W$ boson is chosen.  Besides, photoproductions of multi-jet events, $W+$jets
events, $Z+$jets events, and $t\bar t$ events are all potential backgrounds.
$B$-tagging should be very useful in rejecting a lot of non-$b$ events.

\subsection{Bremsstrahlung Photons}

The cross section using the bremsstrahlung photons is 1.5 -- 5 fb for
$m_H=60-140$~GeV.  Multiplying with the corresponding branching ratios we have
about 3--9 events for the $jjb\bar b$ decay mode, and 1--3 events for the
$\ell\nu b\bar b$ decay mode per each year running.  The largest irreducible
background $ep\to \bar t bX$ has a production rate of 2~pb for $m_t=150$~GeV
\cite{baur}, and $ep\to t\bar tX$ background is much smaller in the order of
0.1~pb \cite{baur}.  In these backgrounds the $m(b\bar b)$ forms a continuum
but in our signal the $m(b\bar b)$  centers at $m_H$.
So by binning on the $m(b\bar b)$,
these backgrounds could be reduced to a 5\% level.  The $ep\to
t\bar tX$  background is reduced to the level of the signal, but $ep\to \bar
tbX$ is still about two order of magnitudes larger.  Further reduction can
be made if we restrict on the invariant mass $m(bW)\neq m_t$.  However, it is
unlikely that this constraint can bring the $\bar t bX$ background down to the
level of the signal due to the smearing of the
momenta of all the particles, and
also it is not applicable for the $\ell\nu b\bar b$ decay mode.
The $ep\to WZX$ background is the same order as the signal in terms of
coupling constant, and should therefore be under control as long as the
$m_H$ is not close to $m_Z$.
Overall, the $ep\to \bar t bX$ remains a major obstacle to the observation
of the $WH$ signal, but the most important is the very small signal rate
that makes
the search via $WH\to jjb\bar b$ or $\ell\nu b\bar b$ almost impossible.

\subsection{Laser Backscattered Photons}

The situation for the laser photons is different because it is in principle
pure $\gamma p$ collision.   The cross section of the $WH$ signal ranges from
23 to 63 fb.  Assuming a yearly luminosity of 3 fb$^{-1}$, we have about
40--110 events for the $jjb\bar b$ decay mode or 13--35 events for the
$\ell\nu b\bar b$ decay mode.  The $t\bar bX$ and $\bar tbX$ backgrounds
only come from the subprocesses $q\gamma\to q'W^* \gamma \to
q't\bar b$ and $q'\bar tb$, respectively.
Unlike the case of bremsstrahlung photons, in which the $\bar tbX$
mainly comes from the subprocess $eg\to \nu W^*g \to \bar tb\nu$, the $t\bar
bX$ and $\bar tbX$ productions using the laser photons should then be of the
same order as the $WH$ signal.  It should also be true for the $WZX$
production via the subprocess $q\gamma \to q' W^\pm Z$.  These backgrounds,
however, have not been calculated.
The $WH$ signal search in $\gamma p$ collision under the presence
of $\bar tbX$, $t\bar bX$, and $WZX$ backgrounds is very similar to the
$WH$ search in the $\gamma e$ collision \cite{WH}.  According to
Ref.~\cite{WH}, the $W^-H$ signal search using the $jjb\bar b$ mode
in the $e^-\gamma$ collision is viable under the presence of $\bar tb\nu$,
$W^-Z\nu$, and $WWe^-$ backgrounds, with or without considering $B$-tagging,
and provided that all the signal lies within $|m(b\bar b)-m_H| <10$~GeV.
Likewise, the discovery of the $WH$ signal in $\gamma p$ collision should
likely be viable, provided that the $m(b\bar b)$ measurement has a resolution
better than $m_H\pm 10$~GeV.

Since the $\gamma p \to \bar t bX$ and $t\bar bX$ are of the same order as
the signal, the binning on the $m(b\bar b)$, say $|m(b\bar b)-m_H|<10$~GeV,
can reduce these backgrounds to a manageable level.  Similarly, $\gamma p\to
t\bar tX$ can be brought down to a level much smaller than the $WH$ signal.
The $\gamma p\to WZX$ background is under control  as long as $m_H$ is not
close to $m_Z$.  When $m_H$ is in the vicinity of $m_Z$, the
discovery of the signal is made possible by a precise calculation of the
normalization of the $Z$-peak, plus there should be
sufficient number of signal events under the $Z$-peak.
Also efficient $B$-tagging is necessary to get
rid of the $W$+jets, $Z$+jets, and multi-jets backgrounds.
A full analysis will be presented elsewhere \cite{new}.

\section{Conclusions}

We have presented  total cross sections of the associated photoproduction
of a Higgs boson with a $W$ boson via $\gamma q \to WH q'$ at electron-proton
colliders with the photon beam obtained by electron bremsstrahlung and by the
laser backscattering method.
At the LEP$\times$LHC ($\sqrt{s}=1.36$ TeV) under consideration, the search for
an IMH boson via the photoproduction of $WH\to \ell\nu b\bar b$ or $jjb\bar b$
with bremsstrahlung photons is almost impossible due to the very small
signal rate.  But with the laser backscattered photon beam there are a
sizeable number of signal events.  Providing that the invariant mass $m(b\bar
b)$ resolution can be made better than $m_H\pm 10$~GeV and the
$B$-identification
is good, the search using $\ell\nu b\bar b$ or $jjb\bar b$ modes should be
viable.  But a thorough analysis taking into account of the full background is
needed to confirm this.  Finally, the laser backscattering option
to perform pure $\gamma p$ collision is favorable in the future $ep$
colliders.

\acknowledgements
This work was supported by the U.~S. Department of Energy, Division of
High Energy Physics, under Grant DE-FG02-91-ER40684.


\begin{table}
\caption{\label{table}
Production cross sections in fb for $\gamma p\to W^\pm HX$ at
$\sqrt{s_{ep}}=1.36$~TeV, using (i) the bremsstrahlung photons, and (ii) laser
backscattered photons, with $m_H=60,80,100,120$, and 140 GeV.  The HMRS(B) is
used for the proton structure function.
}
\begin{tabular}{ccc}
\underline{$m_H$}  & \underline{(i) Bremsstrahlung photons} & \underline{(ii)
Laser  photons} \\
60                 &   4.9    &  63 \\
80                 &   3.4    &  47 \\
100                &   2.5    &  37 \\
120                &   1.9    &  29 \\
140                &   1.5    &  23 \\
\end{tabular}
\end{table}

%
\newpage
\figure{\label{feyn}
Contributing Feynman diagrams for the process $d(p_1)\;\gamma(p_2)\to u(q_1)\;
 W^-(k_1)\;H(k_2)$ in the general $R_\xi$ gauge.
}

\figure{\label{cross}
Total cross sections for the photoproduction of $ep\to WHX$ versus the
center-of-mass energies of the $ep$ system using the WWA for the
bremsstrahlung photons.
}

\figure{\label{cross-l}
Total cross sections for the photoproduction of $ep\to WHX$ versus the
center-of-mass energies of the $ep$ system using the luminosity function
of Eqs.~(\ref{f-lasera}) and (\ref{f-laserb}) for the laser backscattered
 photons.
}


\begin{references}
%
\bibitem{mhlep}E.~Gross and P.~Yepes, Int.\ J.\ Mod.\ Phys. {\bf A8}, 407
(1993).
%
\bibitem{ZZWW}see, {\it e.g.}, SDC Technical Design Report, SDC-92-201.
%
\bibitem{imh}V.~Barger, K.~Cheung, B.~Kniehl and R.~J.~N.~Phillips,
Phys.\ Rev. {\bf D46}, 3725 (1992);
V.~Barger, K.~Cheung, A.~Djoudi, B.~Kniehl, P.~Zerwas, report number
MAD/PH/749.
%
\bibitem{teln}V.~Telnov, Nucl.\ Instr.\ \& Methods {\bf A294}, 72
(1990);
I.~Ginzburg, G.~Kotkin, V.~Serbo and V.~Telnov, Nucl.\ Instr.\ \&
Methods {\bf 205}, 47 (1983); {\it idem} {\bf 219}, 5 (1984).
%
\bibitem{bowser}D.~Bowser-Chao and K.~Cheung, Phys. Rev. {\bf D48}, 89 (1993).
%
\bibitem{WH}Kingman Cheung, Phys. Rev. {\bf D48}, 1035 (1993).
%
\bibitem{zepp}G.~Grindhammer {\it et al.}, in the Proceedings of the
Large Hadron Collider Workshop, Aachen, Germany, 1990.
%
\bibitem{data}K. Hikasa {\it et al.}, Particle Data Book, Phys. Rev. {\bf
D45}, no. 11-II (1992).
%
\bibitem{photo}see for example, G\"{u}nter Wolf, DESY Report number DESY
92-190.
%
\bibitem{HMRS} P.~N.~Harriman, A.~D.~Martin, R.~G.~Roberts, and
W.~J.~Stirling, Phys.\ Rev.\ {\bf D42}, 798 (1990).
%
\bibitem{baur}U. Baur and J.J. Van der Bjj, Nucl. Phys. {\bf B304}, 451
(1988).
%
\bibitem{new}K. Cheung, in preparation.
%
\end{references}
\end{document}